\documentclass{article} 
\usepackage{booktabs,pseudocode,amsmath,amsthm,amssymb,sgame,relsize}
\usepackage{tikz} 
\usetikzlibrary{shapes,arrows}
\usepackage[framemethod=tikz]{mdframed}
\usetikzlibrary{shadows}
\usepackage{paralist}
\usepackage{enumitem}
\setlist{topsep=0pt, leftmargin=*}

\newmdenv[shadow=true,shadowcolor=black,font=\sffamily,rightmargin=3pt]{shadedbox}

\newtheorem{challenge}{Challenge} 

\begin{document}

\title{Models we Can Trust: Toward a Systematic Discipline of (Agent-Based) Model Interpretation and Validation \\ \small{ [Preliminary Version of AAMAS'21 Blue Sky Track Paper]}}

\author{Gabriel Istrate\\
West University of Timi\c{s}oara and the e-Austria Research Institute\\
 Timi\c{s}oara, Romania, email: gabrielistrate@acm.org}

\maketitle

\begin{abstract}  
We advocate the development of a discipline of interacting with and extracting information from models, both mathematical (e.g. game-theoretic ones) and computational (e.g. agent-based models).
We outline some directions for the development of a such a discipline:
\begin{description} 
\item[-] the development of logical frameworks for the systematic formal specification of stylized facts and social mechanisms in (mathematical and computational) social science. Such frameworks would bring to attention new issues, such as \emph{phase transitions}, i.e. dramatical changes in the validity of the stylized facts beyond some critical values in parameter space. We argue that such statements are useful for those logical frameworks describing properties of ABM.
\item[-] the adaptation of tools from the theory of reactive systems (such as bisimulation) to obtain practically relevant notions of two systems "having the same behavior".
\item[-] the systematic development of an adversarial theory of model perturbations, that investigates the robustness of conclusions derived from models of social behavior to variations in several features of the social dynamics. These may include: activation order, the underlying social network, individual agent behavior. 
\end{description} 
\end{abstract}

\textbf{Keywords} agent-based simulation; game-theoretic models; robustness; logical frameworks; 
adversarial perturbations; bisimulation.  

\maketitle

\section{Introduction}

Despite the recent surge in experimental studies of human behavior induced by the availability of (mostly online) social data, a large percentage of work in the mathematical and computational social sciences is still devoted to \emph{theorizing}, that is building \emph{models} of social phenomena, rather than analyzing social data. Whether mathematical or computational agent-based ones, 
a dizzying variety of models is proposed and analyzed in the scientific literature. 

And yet, controversy (if not outright dissatisfaction) about the status and true meaning of such models, and of the modeling process itself, is prevalent throughout the social sciences. An example is the vivid debate on the role of (mathematical) models in economics \cite{sugden2009credible}. Economic models can be interpreted as "credible worlds" \cite{sugden2000credible}, analogies \cite{gilboa2014economic}, (thought) experiments \cite{maki2005models}, parables/fables \cite{cartwright2010models}, intermediate byproducts of robustness analysis \cite{kuorikoski2010economic}, or ludic devices similar to children's toys \cite{toon2012models}. In any case, the discussion about the \emph{robustness of scientific models}, currently taking place in the Philosophy of Science literature \cite{weisberg2006robustness}, is highly relevant. 

A similar debate takes place in the social simulation literature. The critical research problem is that of {\em verifying} and {\em validating} agent-based models (ABM) \cite{axelrod1997advancing,beisbart2019computer} (in short, the {\em V \& V problem}). Theoretical frameworks have been proposed that attempt to deal with this issue, such as the {\em generative approach to 
social simulations} \cite{epstein-generative,epstein-generative-book}, model calibration, {\em docking/alignment} \cite{docking,edmonds2003replication}, {\em replication} \cite{wilensky2007}, model-to-model analysis \cite{hales2003model}, etc.  But there is no consensus on what verification and validation mean (see also \cite{kuumlppers2005,boero2005,windrum2007,moss2008}). 

There are multiple reasons that make the V\&V problem  important and difficult. A first reason is {\em scale}: whereas  Schelling \cite{schelling-segregation} could conceive his celebrated segregation model using pen and paper only, recent simulations models and projects aim to reach global dimensions  \cite{transims,EG+04,bishop2011futurict,barrett2008episimdemics,barrett2013planning}. A second reason has to deal with the potential {\em  social consequences}: social simulations increasingly serve as consultants to (and implicitly affect) public policy \cite{martinez2006modeling,nih-flu2010,marathe2013computational}. A dramatic illustration of this fact in the context of the global pandemic of 2020 has been the controversy around the recommendations of the Imperial College epidemiological model \cite{ferguson2020report}. This has led to significant discussion in the social simulation community, illustrated e.g. by the programmatic article \cite{squazzoni2020computational} and the subsequent comments (e.g. \cite{de2020no,steinmann2020don,edmondsbasic,chattoepolicy,gostoli2020sound,edmonds2020good}). A final reason that makes the V\&V question difficult is the very nature of simulation models, incomplete abstractions of reality, subject to complex behavior \cite{social-emergence} that often involves multiple types of emergence \cite{gilbert-varieties-emergence}. 

It has been noted \cite{lorscheid2019cases} that the proposals put forward in the ABM literature often have an  ad-hoc nature, and that a more systematic theory is needed. \emph{The goal of this paper is to advocate the use of logic and formal methods as useful tools for the systematic development of such theories}.  We discuss a number of ways in which this may happen, and outline several research challenges associated with our proposals. The distinctive feature of the kind of frameworks we advocate is that \textbf{they require a highly unusual combination of two areas with very different languages: logic and formal methods \cite{clarke1999peled}, on one hand, sociological theory \cite{coleman1994foundations}, on the other.}  Importantly, the logical frameworks we envision should actively seek to avoid becoming what Edmonds \cite{edmonds2004formal} called the ``philosophical approach'' to logic. Instead, they should attempt to formalize genuine aspects of social theory (e.g. organizational logic, see e.g. \cite{grossi2005foundations,dignum2012logic}), help with addressing issues related to V\&V, and serve as "middleware" to the agent-based simulations, helping in advancing conclusions that are robust and believable. 

\section{Formal tools for social theory}
\label{seq:logic}

Is there any role for logical formalizations in describing and analyzing social dynamics, in ABM in particular ? This is a question that seems to have been asked so many times, with so many different interpretations in mind that a complete survey of this literature would not be particularly enlightening. Early on, Elster \cite{elster1978logic} argued that ``logical theory can be applied not only in the formalization of knowledge already obtained by other means, but that logic can enter in the creative and constructive phase of scientific work'' (op.cit. pp. 1). He explored the role of {\em quantified modal logic} in describing social reality, with a particular focus towards developing  his method as an alternative to Hegelian dialectics. 
Closer to present Hannan \cite{hannan2007logics} (see also \cite{peli1994logical, peli2000back}) proposed a rational reconstruction  of social theory (organization science in particular) using techniques based on first-order predicate logic. Logical methods are, of course, well-established in economics. To give just one example, the so-called {\em interactive epistemology} program \cite{aumann2000collected} is by now a classical part of theoretical economics, and a key ingredient of a recent proposal for a common foundation of all social sciences \cite{gintis2009bounds}. 

The use of logic-based methods would certainly \textbf{not} be controversial to a large part of the AAMAS audience: In fact, one could justifiably ask what is novel in such a proposal. After all, formal methods based on temporal logic are a particularly significant success story - techniques such as {\em model checking} \cite{clarke1999peled} and {\em runtime verification} \cite{barringer2004rule} lie behind eliminating errors in designing computer circuits, in writing software for technological artifacts (from remote controls and mobile devices to airplanes) or the Mars Rover \cite{brat2004experimental}. Logical methods are widely used in in the area of multiagent systems \cite{wooldridge2000reasoning,wooldridge2002introduction,shoham2009multiagent}. Model checking techniques are useful in the verification of software agents \cite{wooldridge2002model,lomuscio2006model,lomuscio2009mcmas} and auctions \cite{tadjouddine2008abstractions}.   

Yet, \textbf{the above optimism seems not to be shared by the practicing social simulation community.} The mentioned advances in {\em software agents} do not necessarily translate into corresponding advances on simulating {\em social agents} \cite{dignum2020agents}. The techniques developed in the former literature rely too little on existing sociological knowledge, and address to an insufficient degree the concerns of social scientists. Unsurprisingly, they have been criticized (Edmonds \cite{edmonds2004formal}, see also \cite{fasli2004formal,dignum2004use,gaudousocial}) as "not useful given the state of MAS" and "not [...] useful in either understanding or building MAS".

We believe that logical methods can indeed help in increasing the reliability of conclusions derived from social simulations. However, to be useful, \textbf{such logics have to be tailored to the needs of the social scientist, not defined as an object of intrinsic mathematical interest}, and have properties that make them useful: 
\begin{description} 
\item[-] the logics to be developed should be expressive enough to help formalize not only game-theoretic aspects of social theories (see e.g. \cite{parikh2002social,pauly2003logic,van2014logic,perea2012epistemic}) but also a variety of aspects of \emph{sociological theory} \cite{coleman1994foundations,hedstrom2009oxford}. We give in the sequel two examples of concepts that we would like to see formalized: \emph{stylized facts} and \emph{social mechanisms}. 
\item[-] the study of logical frameworks we propose \textbf{should be driven by considerations related to their implementation in (and applications to) ABM}. Their primary goal should \textbf{not} be that of enabling deductive reasoning about social phenomena. Instead, they should be used to \textbf{formally specify} the observed social facts, in a way that enables \emph{the construction of  automated "monitors" serving as "middleware" between the social simulation and the decision support level by recognizing (and signaling) the emergence of the given fact in a given simulation run.} Our proposal is naturally related to the recent call for the development of live simulations \cite{swarup2020live},  i.e. continuously feeding a simulation model with real-world data. In contrast, however, our proposal is related to (automatically) \emph{extracting data from the simulation}, and using it to understand in a more systematic fashion the unraveling of the social dynamics. 
\item[-] \textbf{it is not that important whether deciding implication in the new logics  is tractable}  (we can just run the simulation to see if a certain fact becomes true). However the \emph{model checking} problem  (given a description of the state of the world, is a stylized fact true in it ?) should have efficient algorithms (see also \cite{halpern1991model}). 
\item[-] one problem of significant importance is the \emph{monitoring question} for a logical formula $\phi$: given a sequence of "states of the world" $W_{i}$ (corresponding to a simulation run) and a statement $\phi$, how do we efficiently detect that $\phi$ becomes true at some point in $(W_{i})$ ? This is a question pertaining to \emph{runtime verification} \cite{bartocci2018introduction}, so we should use the inspiration from this literature but, given the rooting of the logical frameworks in social theory, it is likely that a simple adaptation of existing logics will not be enough. 
\item[-] several new research topics, motivated by our vision of studying "robust" stylized facts observed from simulation runs, may gain preeminence. We give an example: the study  of "continuity" properties of parameterized families of logical statements, as we vary the parameters of a given model. The opposite scenario, that of emergence of \emph{critical points (phase transitions)} in the properties of social systems (and in their logical description) is also interesting. 
\end{description} 

\subsection{Parameterized logics of stylized facts: "continuous statements" and "phase transitions"} 
\label{subsec:stylized} 
A first application domain for the logics we envision is the formal specification of \emph{stylized facts}. There is no agreement what a stylized fact is (however, see \cite{meyer2019use}, as well as \cite{floridi2008method} for some relevant philosophical work). To advance a working definition, according to the former paper, at least in microeconomics, ''stylized facts are currently understood as broad, but \textbf{robust enough} statistical properties pertaining to a certain economic phenomenon". 

The requirement that stylized facts are robust is crucial in deciding what is and what is not a \emph{useful} stylized fact: consider e.g. the following trivial baseline scenario (only important as a pedagogical example): 
Each of $n$ agents may be in one of two states, $A$ and $B$. Each agent prefers state $A$ to $B$. Agents are scheduled at random; when scheduled,  each agent changes its state according to the {\em best-response dynamics}, moving to the state that gives it the highest utility. Hence, when scheduled they will turn to state $A$ (and subsequently stay that way, even if scheduled again). 

An obvious conclusion about the dynamics, and a candidate stylized fact, could be the following: \emph{eventually every agent will play strategy $A$}. This is not, however, a \emph{robust} stylized fact. This can be seen by parameterizing the baseline model and modifying agent behavior: we will assume a single parameter $\epsilon \geq 0$. When scheduled, an agent will choose $A$ with probability $1-\epsilon$ and  $B$ with probability $\epsilon$. The baseline model corresponds to the case $\epsilon = 0$. 

It is easy to see that the proposed stylized fact ceases to be true for $\epsilon > 0$, i.e. as soon as we move away from the baseline model. In other words, \emph{the conclusion that \textbf{every} agent eventually holds state $A$ is not robust to even the slightest variation in agents' choice probability}, hence \textbf{it cannot be considered as a (robust) stylized fact}. A more robust formulation is one that claims that agents' state converges to a \emph{stationary distribution} with each agent independently being in $A$ w.p. $1-\epsilon$ and $B$ w.p. $\epsilon$. 
Note that:  
\begin{description} 
\item[-] to formalize the robust version of this stylized fact we don't deal anymore with individual statements, but with \emph{parameterized families} of logical statements. They encode a (single) social fact, expressed slightly differently across variations of the model. 
\item[-] in a very well-defined intuitive sense \textbf{the baseline fact (all agents eventually adopt state $A$) is "the limit", as $\epsilon \rightarrow 0$ of the corresponding parameterized statements for $\epsilon > 0$.} Existing logical frameworks cannot, however, deal with such examples: while 
probabilistic/continuous logical frameworks (and their model checking) exist and might be useful in ABM \cite{kwiatkowska2011prism,platzer2012logics}, and parametricity is important in such settings \cite{ancona2017parametric}, \emph{at the metalevel} logic is still largely a discrete framework, with no concept of "distance between statements", or "continuous limits of statements"
\item[-] in other scenarios the continuous behavior of stylized facts is no longer true. Instead, \emph{social systems display \textbf{phase transitions}}: abrupt changes in the validity of certain stylized facts beyond some critical value of a given parameter. While the study of phase transitions is a well established topic in Complex Systems and A.I. \cite{cheeseman-kanefsky-taylor,selman1995stochastic}, with phase transitions apearing even in settings relevant to model checking \cite{continuous-discontinuous-journal}, the \emph{logical} study of such "phase transitions" is still a relatively underdeveloped area. An exception is the topic of "zero-one laws" in the theory of random graphs  \cite{spencer2001strange}. There are many social phenomena where such concepts seem relevant. An example is the discussion about \emph{tipping points}. Whether one talks about natural or social phenomena \cite{centola2018experimental} there is a considerable interest in anticipating such tipping points \cite{scheffer2010foreseeing}. In the theory of random graphs \emph{the characterization of monotone properties that have "phase transitions" is fairly well understood:} such properties have a "global" nature, depending crucially on the presence of most of the edges of the network \cite{friedgut:k:sat}. In contrast "local properties", e.g. the existence of a fixed subgraph, lack a phase transition \cite{probabilistic-method}. The nature of logical theories in which one formulates the stylized facts also impacts the detection of tipping points: for instance, the emergence of the giant component in a random graph cannot be "sensed" by first-order logic \cite{shelah1994can}. Finally, \emph{similar results exist in scenarios with a dynamical flavor:} start with an empty graph, add random edges, measuring the time when a certain graph property appears. It may be possible to extend such results to settings relevant to ABM: \end{description} 
\begin{challenge} 
Develop a theory of logical frameworks that admit "parameterized statements", and study "phase transitions" in such statements. Ideally this study would yield algorithmic methods to anticipate "tipping points" in agent-based social simulations. Having such methods would operationalize the discussion about the robustness of stylized facts: to argue whether a given stylized fact holds in reality one could ask whether the parameters of the real world lie in the region of the parameter space where the stylized fact varies continuously.   
\end{challenge} 
\subsection{Formalizing social mechanisms} 

It is not only \emph{(stylized) facts} that are in need of a logical formalization. After all, in a social simulation we are not interested in facts only, but in illuminating the \emph{causal reasons} that lead to their emergence. Often (e.g. in the area of Analytical Sociology \cite{hedstrom2009oxford}) such causal explanations involve  {\em social mechanisms} \cite{hedstrom-dissecting,hedstrom-social,demeulenaere2011analytical}.

There is little consensus what a social mechanism is: Hedstr\"{o}m  (\cite{hedstrom-dissecting} pp. 25) compiles a list of seven definitions (due to Bunge, Craver, Elster, Hedstr\"{o}m and Swedberg, Little and Stinchcombe). Of these seven, the most useful is due to  Machamer (\cite{machamer2000thinking}, also \cite{craver2001role, craver2006mechanistic}). As paraphrased in \cite{hedstrom-dissecting} 
{\em ``mechanisms can be said to consist of {\em entities} (with their properties) and the {\em activities} that these entities engage in, either by themselves or in concert with other entities. These activities bring about change [...]. A social mechanism, as here defined, describes a constellation of entities and activities that are organized such that they regularly bring about a certain type of outcome. We explain a social phenomenon by referring to the social mechanism by which such phenomena are regularly brought about''}. 

Social mechanisms are complemented by other approaches: Hedstr\"{o}m lists {\em covering-law explanations} \cite{hempel1965aspects} and statistical explanations. These alternatives are not mutually exclusive: social mechanisms can, e.g., be sometimes inferred from statistical considerations; they can have themselves stochastic/statistical ingredients. 

In any case, whatever social mechanisms are, they seem to have a complex structure: they 
can appear in {\em families} \cite{schelling1998social}, can 
{\em concatenate} \cite{gambetta19985} and be {\em hierarchically nested} \cite{craver2001role}. 
It seems, therefore, that: 

\begin{description} 
\item[-] Verifying and validating social models (including 
simulation models) needs to address issues pertaining to explanation and causality. 
Statistical testing guidelines pertaining to replication,  
such as those discussed in \cite{axelrod1997advancing}, or generative explanations such as 
those proposed in \cite{epstein-generative,epstein-generative-book} are necessary but not 
sufficient. On the other hand social mechanisms,
 being  in one acception ``interpretations in term of individual behavior of 
a model that abstractly reproduces the phenomenon that needs explaining'' 
\cite{schelling1998social} naturally complete and complement these methods (see also \cite{hamill2010}).

\item[-] The role of social mechanisms in validating social models could be 
informally described as follows: simulations should reproduce known social mechanisms 
that are part of the expert knowledge in the area of concern and, of course, perhaps suggest new 
ones. 
\item[-] In accord with \cite{squazzoni2008micro}, ``formalizing models is a prerequisite to illuminate social mechanisms'' and may help in making this notion precise. As a consequence we propose the following 
\end{description}  

\begin{challenge} 
Give logical formalizations of the various notions of social mechanism in Analytical Sociology, and use these formalizations for the automatic recognition and inference of concrete social mechanisms in ABM runs. 
\end{challenge} 

\subsection{Towards a systematic theory of adversarial model perturbations}
\label{sec:efficient} 

It is clear by now that some form of \emph{robustness analysis} \cite{weisberg2006robustness} is crucial to the verification and validation of social models. The concept has been heavily discussed in the Philosophy of Science literature, and can be applied to both mathematical models (e.g. the \emph{robust Volterra principle} \cite{weisberg2008robust}) and to ABM (see \cite{weisberg2012simulation}).  

In contrast there is relatively little work on approaches to robustness with a practical potential: it is known, for instance, that scheduling order can severely impact the conclusions derived from game-theoretic and related models \cite{huberman:glance, weimer2019agent}: indeed, a rich literature on this topic has developed in the cellular automata community (e.g. \cite{fates1997experimental,boure2012probing,fates2004perturbing}). A more general direction, the \emph{adversarial scheduling} approach put forward in \cite{adversarial-mscs} (see also \cite{istrate2008adversarial,shirazipourazad2012influence, istrate2018stochastic}) advocates the study of mathematical and computational models under generalized models of agent activation, as a way to increase the robustness of conclusions derived from these models. Paraphrasing  \cite{adversarial-mscs}, adversarial scheduling is specified by the 
following principles:

\begin{itemize}
\item[-] {\em Start with a ``base case'' stylized fact $P$,  valid under 
a particular (scheduling) model, often random.} Then \emph{attempt to "break $P$" by creating adversarial schedulers under which $P$ no longer holds true.} 
\item[-] \emph{Analyzing perhaps these examples, identify structural properties of the scheduling order that causally impact the validity of $P$. Use these insights to generalize $P$ "from below" by identifying classes of schedulers (including the random one) under which $P$ is valid.}
\item[-]  \emph{In the process \textbf{we may need to  
reformulate the original statement in a way that makes it hold 
under larger classes of schedulers}, thus making it more 
robust.} 
\end{itemize}

As described above, adversarial scheduling is obviously important in increasing the robustness of conclusions drawn from mathematical models: But could something like this be systematically implemented, and be useful for (logic-based specifications of) social simulations as well ? We believe that the answer is positive, and are going to give a pedagogical example, using the baseline scenario from Section~\ref{subsec:stylized}. Indeed, one can logically describe the candidate stylized fact in temporal logic as $ (\forall i) \lozenge [State(i)=A]$ 
("every agent will eventually hold state $A$"). Is such a statement true under adversarial scheduling ? The answer is clearly no: informally, an adversarial scheduler which never schedules a particular agent $x$ whose state is $B$ will preclude the system from reaching the state "all $A$". In other words, to ensure that the baseline stylized fact remains true under adversarial scheduling we need to require the scheduler to be \emph{fair}. The random scheduler is fair (at least with probability $1-o(1)$, as the number of steps tends to $\infty$). 

Could have we reached the above conclusion about the necessity of fairness in scheduling in a logical framework ? The answer is yes. To do so we need to consider a simple logical description of the \emph{the effect axioms} corresponding to the baseline dynamics: 
\[
Scheduled(i)\rightarrow \square [State(i)=A]
\]
 ("if an agent $i$ is scheduled then globally (from now on) the agent will have state $A$"; we  formulated our axiom this way in order to avoid having to deal in this pedagogical example with the frame problem). Can we  derive the statement $ (\forall i) \lozenge [State(i)=A]$, expressing the baseline stylized fact from the action axiom described above ? The answer is negative: to do so we would also need $(\forall i) \lozenge  Scheduled(i)$ ("eventually every agent is scheduled"). However, \textbf{backward chaining \cite{aima} applied to this 
example would identify the statement expressing scheduler fairness as a necessary condition} for the validity of the stylized fact. This is evidence that adversarial scheduling might be feasible even for ABM, thus we propose the following: 
\begin{challenge} 
Extend the theory of adversarial scheduling to more central models of social dynamics, including ABM. 
\end{challenge} 

Scheduling is not the only aspect of a mathematical or computational model which could be studied from an adversarial perspective. Many other aspects are susceptible of a similar treatment. For instance, in many game-theoretic and agent-based models the underlying dynamics takes place on a social network. One may vary this social network and attempt to understand the robustness of the baseline result to changes in the social network. The same can be done with (adversarial perturbations of) initial conditions. Some results in this direction have recently appeared \cite{adversarial-majority}. 

\begin{challenge} 
Develop a theory of adversarial perturbation of social networks and initial conditions for models of social dynamics. Extend it and apply it to ABM. 
\end{challenge} 
\subsection{When are two models ''the same" ?}

The verification and validation problem is related to the the question in the title of this section: when can we really consider two such models, perhaps with different ontologies (e.g. system dynamics and ABM) as "equivalent" ? Again, there is little agreement what a right answer may be. \cite{edmonds2003replication} argue that it is not enough to "eyeball" the outputs of the two models. One of more interesting attempts at an answer is \cite{weisberg2012simulation} (Chapter 8), where model equivalence is formalized as a "weighted feature-matching" problem.  

The theory of reactive systems \cite{milner1989communication} provides an elegant mathematical notion of system equivalence: in this setting, the equivalence of two reactive systems is formalized by the notion of \emph{bisimulation}.  A seminal theorem due to Hennesy and Milner \cite{hennessy1985algebraic} states that two bisimilar systems satisfy the same statements in a certain modal logic $M$, and conversely. That is, bisimilar systems satisfy the same set of "stylized facts" formalizable in $M$.  As impressive as this result is, there is a wide gulf between such theory and the realities of ABM. 
There are multiple reasons that bisimulation is inadequate for social simulation. The most important one is that \textbf{bisimulation  is too "microscopic":} it requires the fact that \emph{every single move} of one of the system is enabled in the corresponding state of the second system. In contrast, cross-validation of ABM is coarser and often qualitative \cite{moss2005sociology}. \emph{In an ABM we don't mean to reproduce the actions of every agents: it's only \textbf{macro patterns} that we care about.} 

There is some hope, though, that such methods are relevant to the study of ABM after all: recent results \cite{pauly1999game,gutierrez2017nash,wooldridge2019nash}, some even from AAMAS \cite{belardinelli2017bisimulations} have related bisimulation to game-theoretic scenarios. It is thus reasonable to propose 
\begin{challenge}
Develop a theory of (bi)simulation of (social) systems aligned with (an relevant to) the practice of V\& V in ABM. 
\end{challenge} 
\section{Conclusion} 

We believe that logical formalization plays an important rule in assessing (and increasing) the reliability of results in social simulations. We have highlighted a couple of research directions that (if successful) would orient and ground the current discussion on model validity in (computational) social sciences. We don't believe that the directions we outlined are going to completely solve this problem. But, besides the obvious intellectual interest of developing such concepts, they may contribute to turning simulation and modeling in social settings from largely being an art (which still is now) to an engineering discipline.

\bibliographystyle{unsrt} 
\bibliography{/Users/gistrate/Dropbox/texmf/bibtex/bib/bibtheory}  

\end{document}